\documentclass[journal]{IEEEtran}

\usepackage[utf8]{inputenc}
\usepackage[T1]{fontenc}
\usepackage{siunitx}
\usepackage[colorlinks=true, allcolors=blue]{hyperref}
\usepackage{amsmath}
\usepackage{amsfonts}
\usepackage{soul} 
\usepackage{booktabs}

\usepackage{tikz,xcolor}
\definecolor{lime}{HTML}{A6CE39}
\DeclareRobustCommand{\orcidicon}
{
    \begin{tikzpicture}
    \draw[lime, fill=lime] (0,0) circle [radius=0.16] 
    node[white] {{\fontfamily{qag}\selectfont \tiny ID}};    \draw[white, fill=white] (-0.0625,0.095) circle [radius=0.007];    
    \end{tikzpicture}
    \hspace{0mm}}
\foreach \x in {A, ..., Z}{%
    \expandafter\xdef\csname orcid\x\endcsname{\noexpand\href{https://orcid.org/\csname orcidauthor\x\endcsname}{\noexpand\orcidicon}}
}

\begin{document}

\title{Development of VR Teaching System for \\ Engine Dis-assembly}

\author{Zhuochen Xiong
\thanks{Zhuochen Xiong is with the Department of Computer Science and Engineering, Southern University of Science and Technology, Shenzhen, 518055, China (e-mail: 11811806@mail.sustech.edu.cn)
}
}

\markboth{Journal of \LaTeX\ Class Files,~Vol.
}%
{Shell \MakeLowercase{\textit{et al.}}: Bare Demo of IEEEtran.cls for IEEE Journals}

\maketitle

\begin{abstract}
  With the worldwide ravaging of the covid-19 epidemic, the traditional face-to-face education systems have been interrupted frequently. 
It is demanded to develop high-quality online education modalities. 
The webcasting based online classroom is one of the popular education modalities
but suffers from poor teacher-student interactions and and low immersive learning experiences. 
This thesis aims to improve the online education quality by using the virtual reality (VR) technology.
For the purpose of automobile engine education, we develop a VR based engine maintenance learning system. 
The system includes many teaching and learning components in VR enabled by the Unity engine. 
Users can immersively experience the complete engine disassembly process through the wearable VR display and interactive devices. 
The system is designed with an interactive layer, a control layer, and a physical data layer. 
Such a system architecture effectively separates the specific implementations of different domains and improves the R\&D efficiency. 
Once new object models and process profiles are provided,
the proposed system architecture requires no modification of codes for changed learning objects and processes.
The efficiency and effictiveness of the proposed method are verfied by various experiments.
The developed techniques can be useful for many other applications.

\end{abstract}


%
\IEEEpeerreviewmaketitle

\section{Introduction}
According to relevant statistics from the Ministry of Education, there will be about 8.74 million college graduates in 2020. The number will rapidly grow to 9.03 million in 2021, increasing 350,000 year-on-year while also reaching a new historical high for the number of college graduates in China. The increasing number of college graduates does not match the demand for jobs. The employment of fresh graduates from colleges and universities is like a thousand horses breaking through a single wooden bridge, so many college graduates choose to study in graduate schools, abroad, and other forms of temporary employment pressure.

In stark contrast, the employment rate in vocational colleges is increasing year after year. This shows that the demand for talents with special vocational skills is increasing day by day. This is also due to the fact that the evaluation system of talents is becoming more and more diversified, not just "academic only". In such a context, the state is paying more attention to the diversion of secondary school graduates to guide more of them to aspire to enter vocational institutions to acquire professional skills, which will make them more competitive in the job market. Therefore, the quality of vocational education is essential.

For the past few decades, high school students have been evaluated by society on their ability to get into a good college, while some students entering vocational institutions have disappeared from the public eye and received no social attention. This has also led to a lack of attention to vocational education and relative backwardness in terms of teachers and teaching resources. This is especially true in areas that require teaching practice, such as automotive repair, electronic device repair, and complex instrument manipulation. Because of the lack of input, it is often difficult for students to have sufficient access to relevant practices during the learning stage. In particular, schools may choose not to teach in areas that have safety risks because they do not have sufficient resources and do not want to take responsibility for safety. At the same time, because of the lack of teachers' resources, it is difficult for the faculty of vocational colleges to meet the needs of a large number of students. As a result, the teaching quality of theoretical courses also varies, which eventually makes it difficult for students in vocational colleges to master the complete knowledge-theoretical system and practical skills.

\subsection{Motivation}

In the post-epidemic era, the face-to-face classroom format is difficult to achieve. For the sake of epidemic prevention and control and the safety of teachers' and students' lives, it is essential to explore distance learning formats that can restore the quality of classroom teaching to the greatest extent possible. In recent years, digital offline teaching forms have emerged, and standard forms such as online video classes have covered almost every stage of education. However, the current digital teaching scene still has problems such as poor interaction and poor immersive experience for teachers and students, leading to poor teaching quality and learning experience than offline classrooms.

There is an urgent need to introduce \textit{metaverse-based} virtual teaching and learning scenarios in this context \cite{lan2021learning1,lan2021learning2}. Participants will have a better sense of presence and interactive experience when using a \textit{metaverse-based} virtual teaching system. These advantages create a surreal virtual integration teaching environment conducive to stimulating learners' curiosity, imagination, and creativity. Further, this is the key that can enhance learner engagement.

Based on the above status, our team developed an engine disassembly system based on virtual reality technology in cooperation with \textit{FXB CO., LTD.} The system is adapted to mainstream VR devices. The teaching process and teaching objects are stored in a parameterized form in the configuration file, which is highly scalable and migratable.

\section{Related Work}
\label{sec:related_work}
Because of insufficient human and material resources in teaching, many schools have started to use systematic teaching software to assist in teaching. Virtual reality-based simulation teaching software has become increasingly popular in recent years and has even been widely used in some western countries.

Currently, the application scenarios of virtual reality teaching systems in China are still minimal and tend to focus on some disciplines with high practical costs. One of the more typical scenarios is anatomy. In some organ anatomy scenarios, experimental materials are usually more precious and difficult for most students to perform hands-on. Students can have a more realistic experience in the virtual reality scenario and significantly reduce the related material costs.

In addition, the advantages of virtual simulation teaching are also reflected in the safety and high reproducibility. In some industries with certain operation safety risks \cite{gao2021neat,lan2017development}, using virtual reality to simulate actual operations can reduce the occurrence of safety accidents. Furthermore, for some scenes with the high complexity of teaching, unified teaching software can easily replicate the teaching process and achieve the purpose of unified standards.

\subsection{Unity3D}

Unity3D is a tool platform developed by Unity Technologies that can be used to create any real-time interactive 2D and 3D content. The Unity platform provides a complete set of tools to help developers quickly build projects. We choose this tool as our primary technology stack based on the following characteristics \cite{xie2012research, messaoudi2015dissecting}.
\begin{enumerate}
    \item \textit{Visual development interface.}  As a tool that can be used to build 3D models, an easy-to-understand visual interface is an excellent advantage of Unity. We can complete project settings and script editing in the visual programming interface, and the development efficiency is extremely high.
    \item \textit{High compatibility with 3D models.} Unity3d supports most 3D models and can automatically convert texture materials to U3D format. Its good compatibility helps us quickly import pre-built project models, including static operating room models and high-precision engine models \cite{wang2010new}.
    \item \textit{Multi-platform compatible.} The unity suite supports a one-click compilation of projects into multiple formats, adapting to powerful software platforms that are currently popular. We can achieve platform-independent project development based on this feature.
    \item \textit{Excellent performance.}  The bottom layer of Unity natively supports OpenGL and Direct11, which helps to render high-precision physical models quickly. In virtual reality, high-precision physical models can improve the user's sense of presence.
\end{enumerate}

\subsection{VRTK \& DoozyUI}
In the project, we used VRTK and DoozyUI to implement the teaching system. The dynamic interaction between characters and models in the scene is implemented using VRTK, and the messaging and user interface in the system is managed using DoozyUI.


Virtual Reality Toolkit aims to make building spatial computing solutions in the Unity software fast and easy for beginners as well as experienced developers \cite{chen2017practice, ltd_2016}. 

In our project, VRTK is used to build the interaction behavior of items, which has the following excellent features.

\begin{enumerate}
    \item \textit{Good compatibility.} VRTK was formerly a Steam VR development tool, but with the development of the community, it gradually supports most mainstream VR devices, such as Oculus, GearVR, etc. Good compatibility can help us develop teaching systems adapted to multiple platforms so that teaching institutions using the system do not have to be limited to the purchase models of VR devices.
    \item \textit{Rich pre-built interaction behaviors.}  As a mature VR development SDK, VRTK has a variety of built-in common interactive actions, such as: grabbing, sliding, throwing, helmet movement, etc. For helmet movement this kind of action, VRTK has a variety of built-in movement methods, such as dash Movement, touch-pad Movement, transient movement, etc. In addition, VRTK produces sample scenes for these common interactions. We can see how these interaction models are implemented in these scenes, including clear C\# code and related settings.
    \item \textit{Efficient keyboard and mouse simulator.} In order to help developers to debug programs more conveniently, VRTK implements a VR debugging simulator that can be completely simulated by keyboard and mouse. This tool helps us develop without equipment, which provides great convenience for our remote collaboration during the epidemic.
    \item \textit{Rich documentation and active community.} As we all know, whether the documentation is clear and the community is active has far-reaching significance for the development of an SDK because it determines how many people will be willing to use and improve its functions. The documentation of VRTK is rich, and the developer can directly view the documentation of related functions by hovering the mouse on the GUI interface, which is highly readable. These documents helped us a lot in the project's early days, allowing us to learn how to use the tool quickly. In addition, VRTK's community activity is also very high. It has obtained thousands of stars in the GitHub community, and many issues are constantly being raised and resolved, helping it to improve its features continuously.
\end{enumerate}


DoozyUI is a well-known UI management framework in the Unity ecosystem. We use it to build a complete user interaction system in our project. Compared with other UI frameworks, DoozyUI has the following advantages \cite{kok2021beginning}.

\begin{enumerate}
    \item \textit{Easy to learn and use.} DoozyUI has a complete graphical user interface and can complete most UI design and construction through GUI. The visual interface is very intuitive when debugging views, reducing the cost of collaboration for developers. For standard interaction models in user interfaces, such as layouts and buttons, DoozyUI provides services to the upper layer in the form of APIs, reducing the cost of development and understanding.
    \item \textit{Components are persistent.} In DoozyUI, each view is treated as a container and can be managed using the built-in container database. Developers can define any UIView and its features in the database. These predefined UIViews can be easily called in any code, and the layout can be adjusted directly in the GUI interface.
    \item \textit{Signal-based event delivery.} The event delivery system is arguably one of the most powerful features of DoozyUI. In the design of DoozyUI, all event operations are called asynchronously in the form of signals. The advantage of unified message events is that different nodes can act as message producers or consumers, and the form of event delivery is abstracted into a typical pattern. A meta-signal in DoozyUI can be anything from a simple value to a reference or action instruction. In our system, meta-signals are widely used to transmit information for system control.
\end{enumerate}

\subsection{Scenic Spheres}

\textit{Scenic Spheres} \cite{bryan2018scenic} is an online game based on virtual reality technology. It uses AR/VR technology to immerse users in infamous sites worldwide, encouraging them to learn about geography through goal-based incentives. In the game scenarios, users are placed in different locations worldwide. In the virtual reality scenario, users must look around for clues, and only after answering as many questions as possible through the clues will new areas open up. The first player to find the flag will be the winner of the game. This incentive-driven learning approach helps motivate students to learn \cite{lynch20top, fowler2013survey}.

This project is also a game project based on unity engine, and C\# is used as the programming language of the project. In virtual reality technology, the biggest challenge is the user's depth perception of the virtual environment \cite{lan2016developmentVR,lan2016developmentUAV}. Because of image resolution limitations, the user may not be able to accurately perceive the depth of the scene, which may cause the user to experience headaches or dizziness. \cite{rolland1995comparison, van2010survey} Therefore, it is necessary to solve this problem as much as possible. From a technical point of view, it is possible to enhance the sense of reality and reduce user discomfort by enhancing the picture resolution. In the field of education, there are more humane solutions. Teachers can reasonably manage the time that students use VR devices and avoid students staying in the virtual environment for too long.

\section{Methodology}
\label{sec:methodology}
\subsection{System Design}
\begin{figure}[!ht]
    \centering
    \includegraphics[width=.45\textwidth]{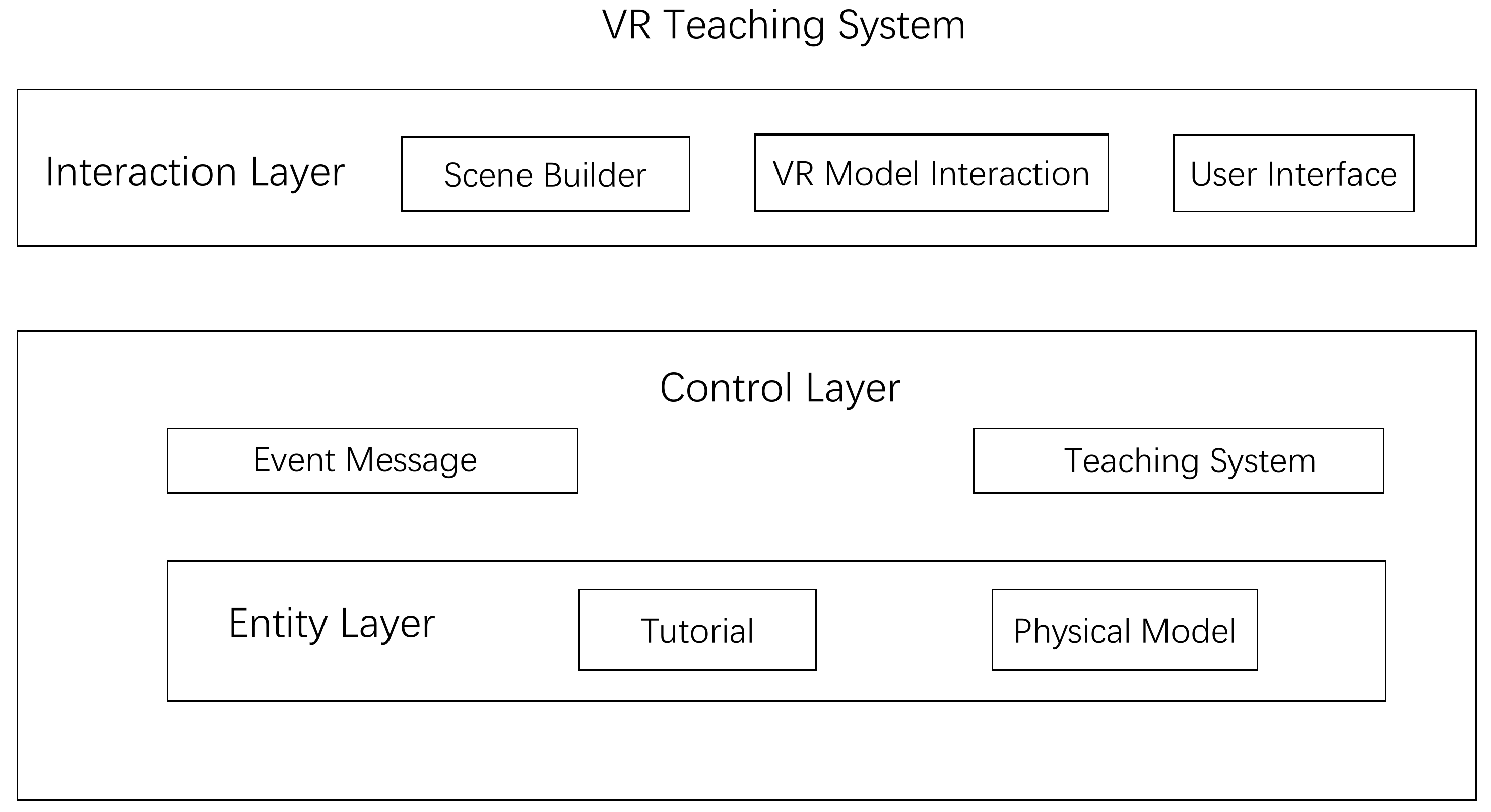}
    \caption{System Design}\label{F:system}
\end{figure}

The primary modeling object of the virtual simulation teaching software we have implemented is the \textit{Buick Verano} engine. Based on the 3D model, the parts are disassembled according to the real disassembly order, and the orderly disassembly process is realized. As a teaching system, besides the basic interactive operations of disassembly and assembly, it also needs to have teaching logic. For example, different parts are divided into 
teaching chapters, and the difficulty is divided into two stages: training and examination. In order to achieve high-performance, scalable simulation software \cite{lan2019simulated,lan2019evolutionary}, we abstract the software architecture into the following form as Figure \ref{F:system}.

Specifically, our teaching software abstracts the behavior interaction between the user and the teaching model into an independent behavior for coding. It then defines a series of rules for this specific behavior. From a functional point of view, we use to \textit{teaching} and \textit{examination} to distinguish user interaction into two independent scenarios. In these two scenarios, we endow the software with different behaviors. The 
teaching mode is more focused on teaching disassembly and assembly skills, so in this mode, we use text tutorials and voice prompts on the way home to guide users to complete the specified actions. Then, the examination mode focuses on inspecting the user's skill mastery. In this mode, we will turn off the voice and text prompts, add a series of scoring rules to score in different dimensions according to user operations, and finally give a user evaluation of operational performance. Further, we can also develop this storage system of user scores from local records to online records so that teachers can count students' learning and improve teaching efficiency.

From the business scenarios described above, it can be seen that the two scenarios of \textit{teaching} and \textit{examination} are the difference between the education business. However, their corresponding underlying logic is almost identical, and they are all the same disassembly or installation process. This means that interaction and teaching logic can be decoupled and placed at different architectural levels, as shown in the Figure  \ref{F:system}.

Architecturally, the teaching system uses a bottom-up design. Through the entity layer, the developer completes the modeling of the demand scenario, builds the overall framework of the entire system, and abstracts the teaching process into a configurable data structure. After completing the construction of the entity layer, we will enter the research and development of the control layer, which subdivides the specific business logic. Divide the teaching process and the specific logic of model interaction into atomic logic, and complete the design of these logic at the code level. After completing these logic designs, we will eventually reassemble the basic atomic logic into complete business logic at the interaction layer according to specific product requirements. Under this architecture, product iteration only needs to modify the control layer in the middle because the basic operation interaction scripts are reusable. When encountering a new teaching scene or facing a new interactive model, the separated entity layer can be individually replaced with a new teaching scene. Different processes can be used to interact with the original view.

Such a design ensures the high scalability of the system and enables the template-based development of the auto-repair immersive teaching software, which significantly improves the development efficiency. Similarly, such a design also helps project team members efficiently divide their labor, splitting the coupled business logic into different parts at the level of code implementation and combining different functions more flexibly. The content of each level will be described in detail below from a business perspective.

\subsection{Interaction Layer}

The interaction layer is the top-level structure in the software system, which includes everything the user can see. From the function division, we can divide it into three modules: scene construction, model interaction, and user interface.

\subsubsection{Scene Builder}

Scene builder is the user's surrounding environment construction in the virtual reality scene. In order to enhance the user's sense of presence, our teaching system sets the scene after the user enters the system as the teaching demonstration room of FXB CO., LTD. This module belongs to the field of 3D modeling and does not involve coding work. In this project, the company provided us with the 3D model of the teaching demonstration room and its corresponding textures and materials. We import the corresponding model into our project in a static file and create an instance through unity3d. These 3D object instances as operating environments do not involve model interaction and are only rendered statically in the scene.
\begin{figure}[htbp]
    \centering
    \includegraphics[width=.45\textwidth]{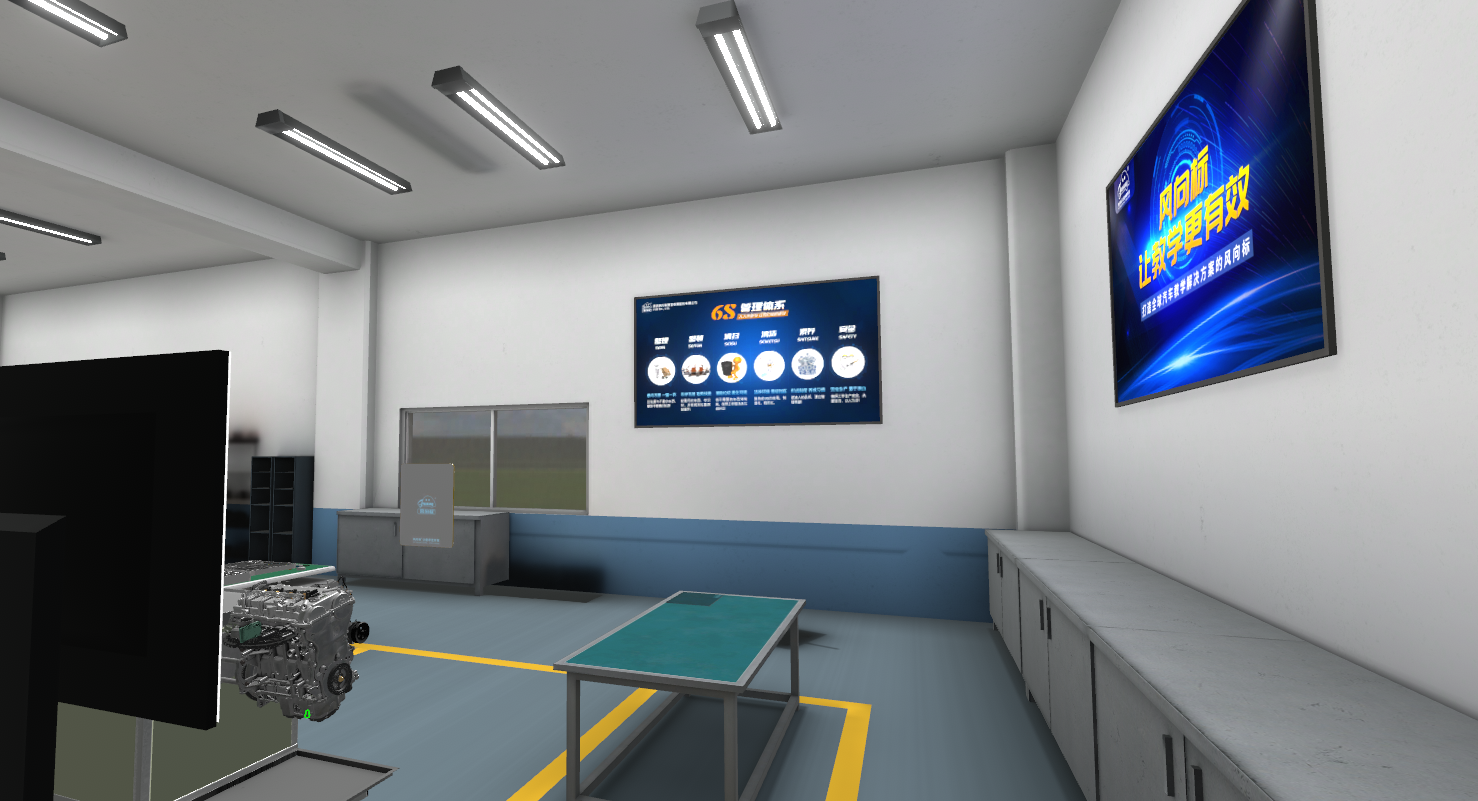}
    \caption{Virtual Teaching Room in FXB CO., LTD.}\label{F:scene}
\end{figure}

\subsubsection{Model Interaction}

Model interaction is the core interaction scene of the system. This module contains two parts of model construction and interaction. For model construction, it is necessary to use the rendering technology provided by Unity3D to give it natural light and shadow effects to show a natural luster in the virtual scene.

The interaction of the model is the core part of the project. The models can be divided into two types according to the interaction properties in this project. One is non-interactive static scenes, such as buildings, floors, shelves, Etc. The other is an interactive dynamic model, a \textit{Buick Verano} engine, and a display and pad that can display content dynamically.
\begin{figure}[htb]
    \centering
    \includegraphics[width=.45\textwidth]{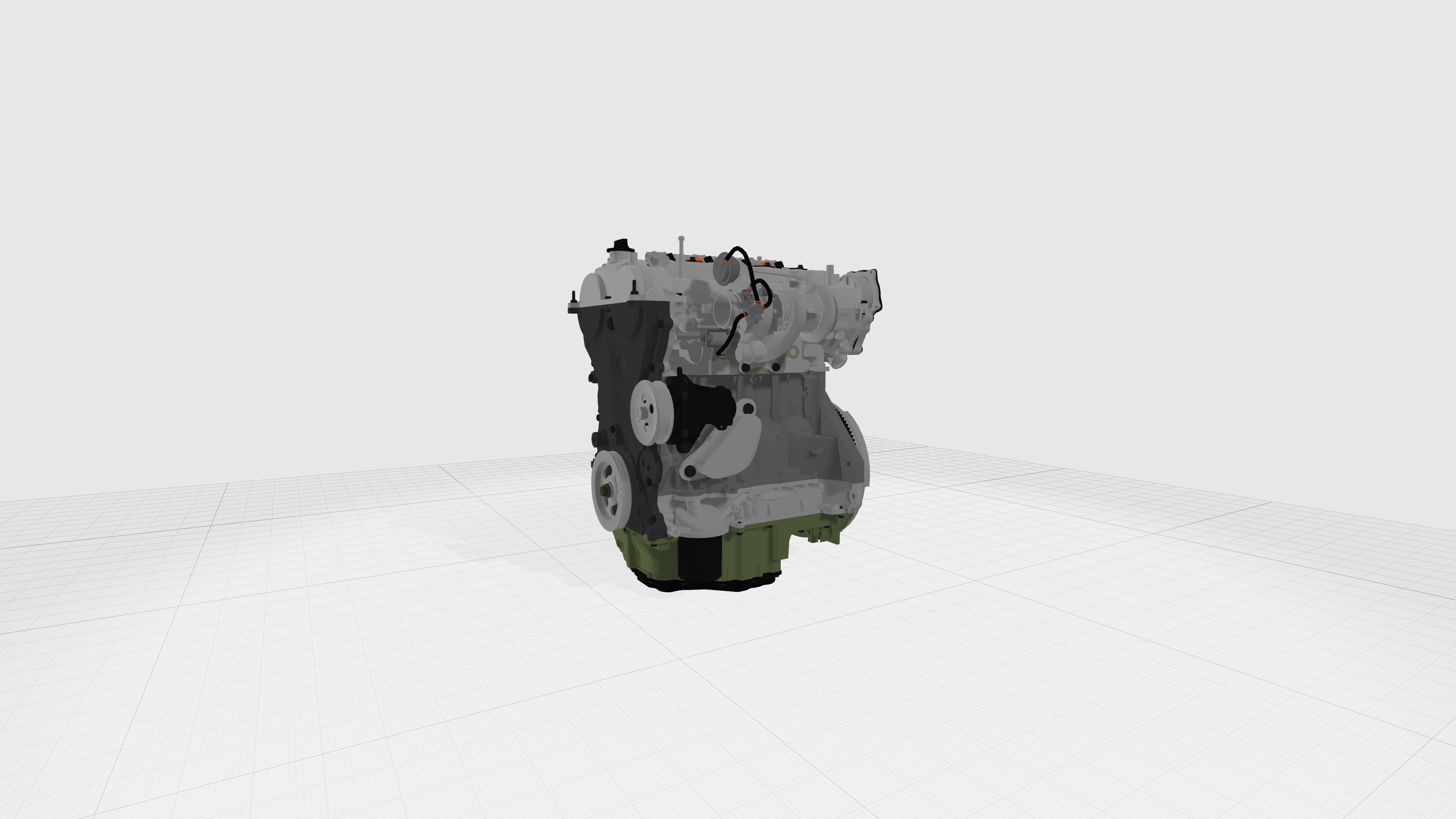}
    \caption{side view}\label{F:verano}
\end{figure}
\begin{figure}[htb]
        \centering
        \includegraphics[width=.45\textwidth]{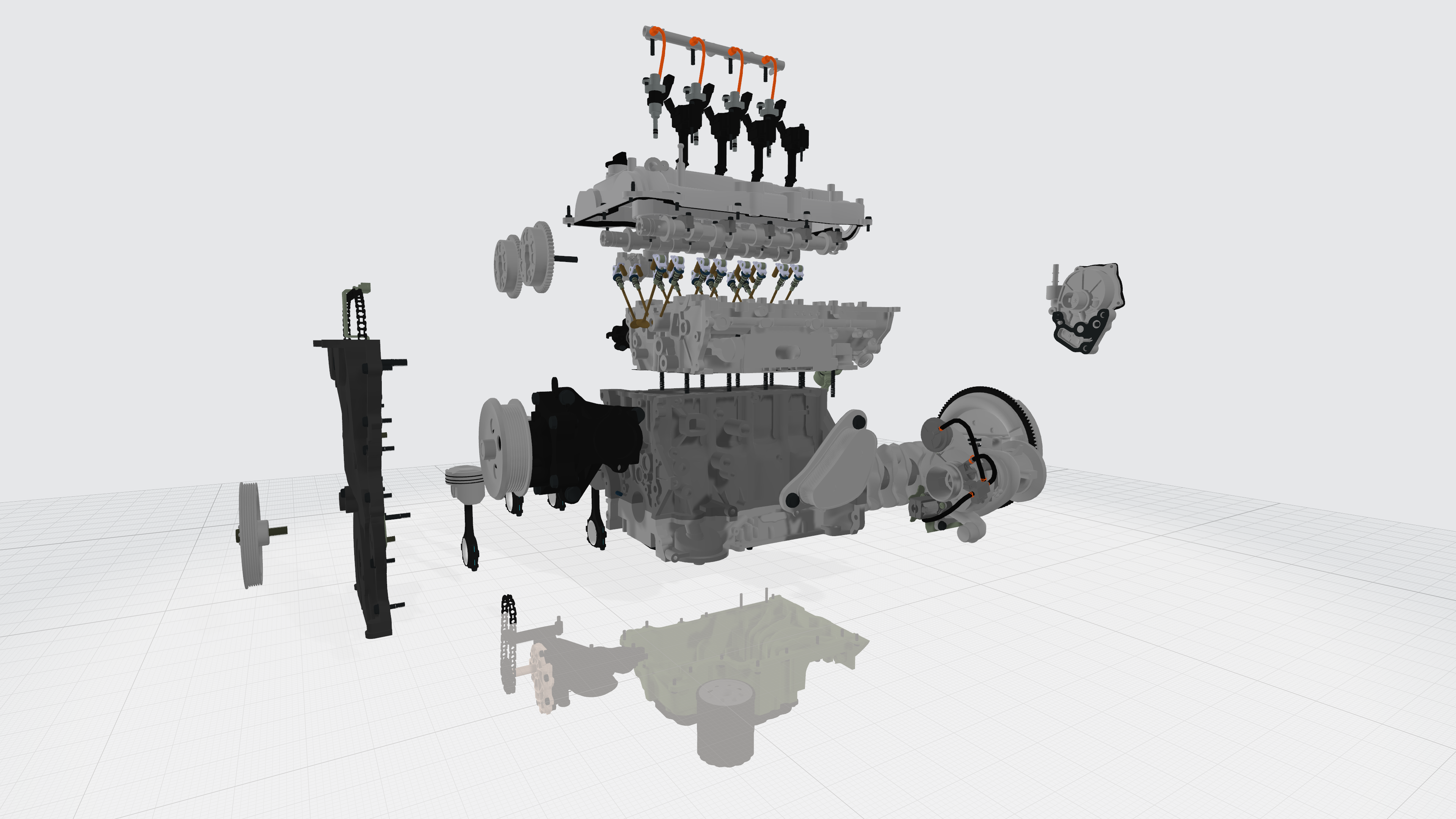}
        \caption{exploded view}\label{F:verano-boom}
\end{figure}

The \textit{Buick Verano} engine is our teaching system's teaching object and the main object of model interaction. For an engine model, we can think of it as a collection of parts. These parts are the direct use objects of our ActionScript.
Parts come in various models and can have hundreds or thousands of shapes. However, for a part, the combined actions with other parts are limited, and these actions can be split into some basic actions: rotate, press, take, hide, Etc. All teaching actions can be broken down into combinations of these actions. The combination of rotation and pressing is the most frequently used because screwing is the most frequent among all actions.

\begin{figure}[htb]
    \centering
    \includegraphics[width=.5\textwidth]{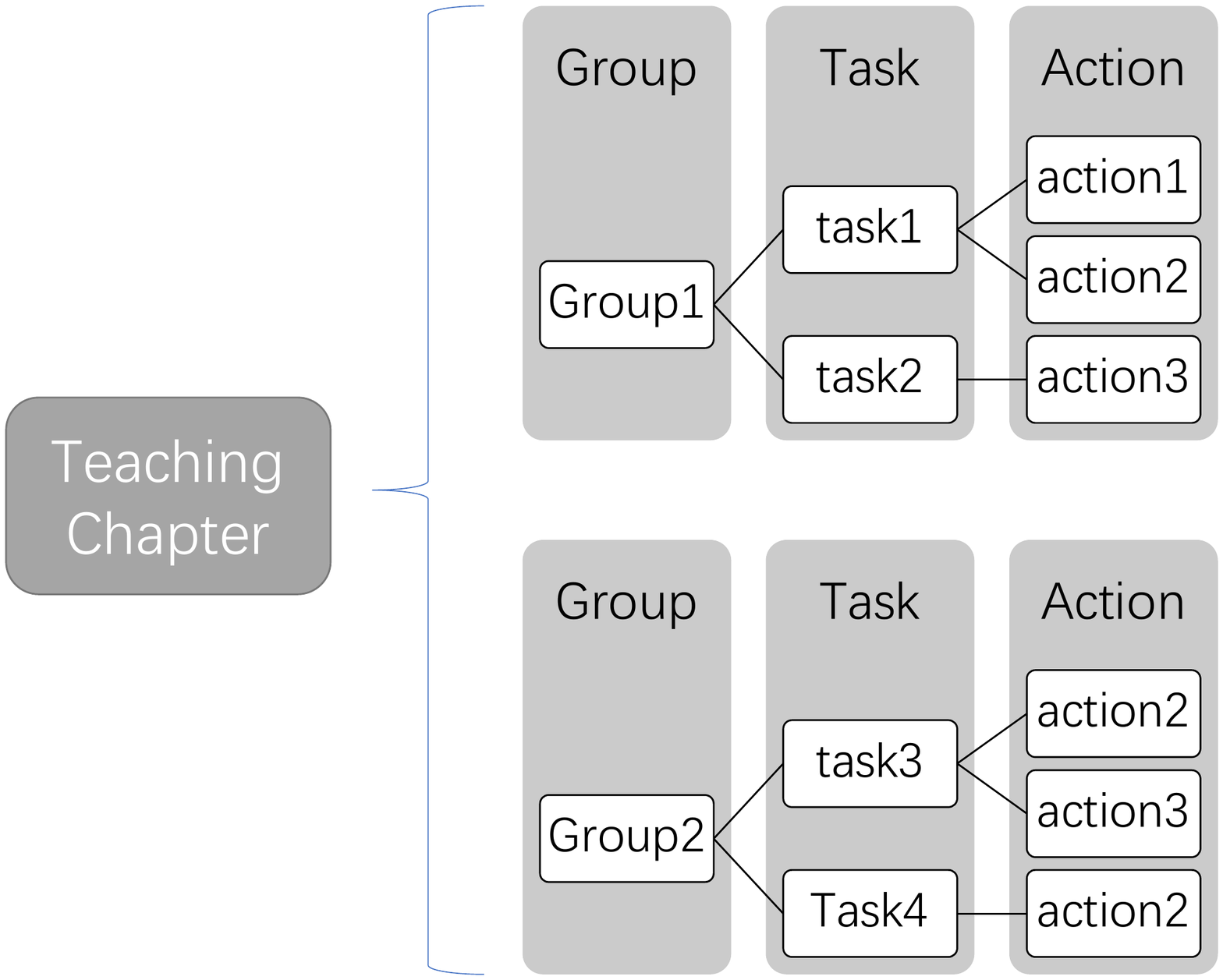}
    \caption{config file structure}\label{F:files}
\end{figure}

Based on the above ideas, we can abstract the concept of action into an atomic capability. For each atomic capability, we use a C\# script to describe the behavior of its object. However, in actual business, each action in the teaching process is not a single atomic action but a combination of some atomic actions. Therefore, we need an efficient way to organize these combinations in an orderly manner, that is, configure the content and order into a configuration file for persistent storage. In storage, we organize it as figure \ref{F:files}.

In the system operation process, we can uniquely identify a currently ongoing task. During the running of the task, we will read the corresponding step information under the task. When the user triggers a step, the atomic action corresponding to the step will be executed, monitoring whether the user completes the action as required.

\subsubsection{User Interface}
The user interface is an integral part of the software system and is the entry point for the user to interact with the system. In our virtual reality engine disassembly system, the user interface can be considered everything the user can interact with within the virtual reality scenario. However, to better abstract the user interface concept, we only define the interaction part that affects the system state as the user interface. The pad in the scene with the screen displaying information satisfies the definition.

In a virtual reality scenario, the user interface can be a combination of a series of views, each containing interactable components (such as buttons, sliders, and selection boxes) and non-interactive static resources (such as images and description text). These views are mainly managed and controlled using the DoozyUI Manager in our design.

\begin{figure}[!htb]
    \centering
    \includegraphics[width=.48\textwidth]{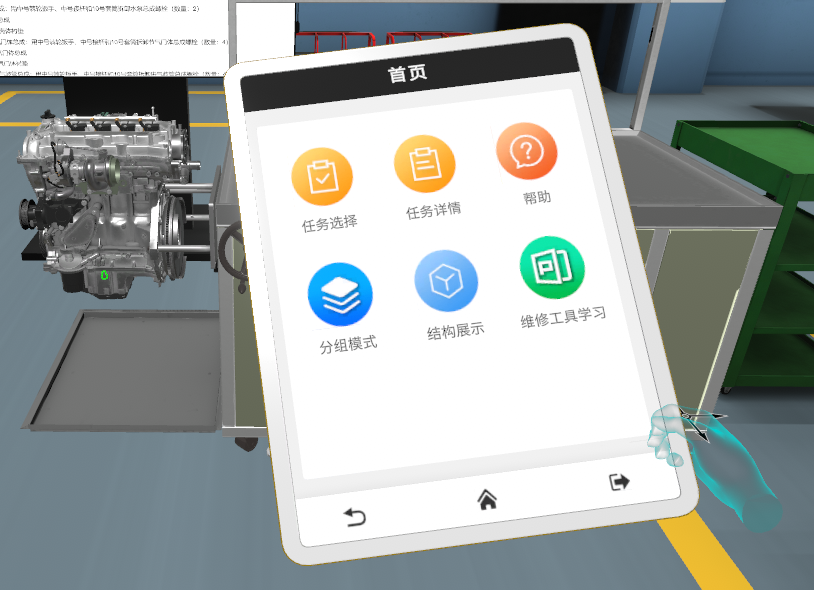}
    \caption{Pad operation in teaching system}\label{F:pad}
\end{figure}

The pad is the main entry point for the user interface in our system. As shown in the figure, we control the system process through the tablet. We provide six entrances on the pad's home page that carry the two main functional modules of teaching and  examination.

We provide three relevant entrances for the teaching function: grouping mode, structure display, and repair tool learning. In the grouping mode, users can choose the chapter they wish to enter. Entering the system in this portal will be recognized as training mode, which does not rate the user's operation and will have voice prompts during the operation on how to proceed to the next step. The remaining two portals do not involve interaction with the model, and both present the user with static resources in the form of tutorials. One of them, Repair Tool Learning, is a video resource that will automatically play an instructional video of the repair tool when this button is clicked.

For the examination function, we provide two entrances, task selection, and task details. The main entrance is task selection, where users can enter the system and select the disassembly or installation task they wish to perform. When performing a disassembly or installation task, the large screen will show the steps currently in progress. If the user needs to see all the steps in this task, they can click on the Task Details button, which triggers the screen to show all the steps of a task.

We manage all views, buttons, and canvases in the pad through DoozyUI. These views can be predefined and stored in DoozyUI's database, and then their corresponding jump relationships are organized in the scene by setting the timing of when each view is shown and hidden.

\subsection{Control Layer}
The view layer is the direct object of user interaction with the system, while the control layer is the inner logic that implements all user interactions at the view layer. This section will elaborate on how the user's interaction actions affect the system state.

\subsubsection{Event Message}

The calling relationship between each module is very complicated in the teaching system. When the user enters a particular chapter to study, the state of the objects in the scene needs to be updated synchronously. Each time the user completes an action, the corresponding action guide and corresponding voice prompt need to be refreshed. In addition, various multimedia teaching resources also require different responses between the iPad and the screen according to the user's interactive behavior.

Because the call relationship is complex, the communication between components will make the system more complicated if the synchronous call method is used. Therefore, we design a component communication method based on message passing, which decouples message notification and action to improve coding efficiency.

The message system is implemented based on the DoozyUI signal system. In DoozyUI, we can define the instance as the sender or receiver of the message, and we just need to override the \textit{SendSignal} method and the \textit{Get} method.

In our system, a message consists of an action type and a target object that needs to be changed. The delivery of messages in the business has the following situations:

\begin{enumerate}
    \item Display the corresponding media content on the large screen according to the user's click behavior on the iPad.
    \item Trigger the update of ongoing task steps according to the user's disassembly and assembly actions
    \item Automatically play the corresponding step prompt audio according to the user's disassembly and assembly behavior
    \item System state changes based on parts collision interaction
\end{enumerate}

Under such an architecture, each component in the system does not need to perceive the actions of other components and only needs to design its behavior when sending and receiving messages. Such a structure helps different components to develop independently. After defining the specifications of the good news, the development can be completed independently by different team members, improving team collaboration efficiency.
\subsubsection{Teaching System}

Teaching is the core function of our system. We have designed three teaching modes: teaching, practice, and examination. In the teaching mode, we provide an automatic voice guidance function. After the user completes a step, the voice guidance of the following action will be played automatically. At the same time, in the teaching mode, users can also view video tutorials to learn the usage of various tools and basic engine disassembly and assembly skills. The practice mode is an advanced version of the teaching mode. There is only brief operation prompts in the practice mode instead of the complete teaching process. In this mode, users have total freedom to practice the practical knowledge they have learned.

Further, we designed an examination mode to assess the user's mastery of a specific task. In examination mode, the weight of each step can be predefined. When a user completes a task, the completion status will be recorded, and the machine will score the user's performance according to the weight of each step. There will be no prompt content in this mode, the user needs to complete the examination independently, and the terminal will record the results.

The teaching system is organized based on the user interface, and the user can freely choose different mode entrances on the Pad in the scene. When the user enters a specific task, the system will recognize which teaching mode the user is currently in and send a message indicating entering the task to the scene. The teaching scene will determine how to refresh the page at this time according to the content of the message, reset the engine state to the position required by the task, and set the prompt content of voice and text to the corresponding teaching mode.

\subsection{Entity Layer}

Entity layer is the storage layer for data. We define our business entities in this layer and store entity information persistently. After decoupling system design and requirement design, we can abstract the concept of requirement entity. In this teaching system, the demand entities that change with the business changes in the system can be roughly divided into two types: tutorial entities and physical model entities.

\subsubsection{Tutorials Entity}

Tutorial entities represent what students learn in our actual needs. In this project, the tutorial for the students is the disassembly and assembly of the \textit{Buick Verano} engine. When the demand scene is changed to the engine disassembly and assembly tutorial of other brands, or even the disassembly and assembly of other items, we need to be able to change the tutorial entity easily. Therefore, our system provides the abstract concept of tutorials in a configurable form.

\subsubsection{Physical Entity}

The physical entity is a collection of models in a scene. In a user interface, the object that the user directly interacts with is the model in the scene. However, these models should not be coupled to the system design, and we want our system to be usable in any simulation scenario. The physical models in the system include static scenes: demo room, console, toolbox, and dynamic models: engine, pad, Etc. These contents make up our physical entities.

We build these entities in a statically modeled fashion. Each of these models is a separate 3D model file. We preset their shape, color, and physical properties and import them into Unity. These model files can also be reused in other projects.

\section{Experimental setup}
\label{sec:setup}
Evaluating our systems is also an essential part of our work. Our system is a teaching system based on virtual reality so that the evaluation can be carried out from the two dimensions of virtual reality experience and teaching experience.

The teaching experience cannot be quantified using specific performance data but must be collected from user experience report statistics after a small-scale pilot to conclude. Therefore, the assessment process in this report does not involve the assessment of the teaching experience. Promoting this project in some vocational schools and recycling the experience report will be critical work in the future \cite{xu2019online,lan2018real}.

In the field of virtual reality, the quality of experience can be assessed from the following dimensions:
\begin{enumerate}
    \item \textit{Immersion.} This is a more specific concept, indicating whether the user's perception of the scene after entering the system is real. This dimension needs to be evaluated by multiple factors, including the clarity of the video, the spatial sense of the audio, and so on.
    \item \textit{The viewing experience.} Whether the information received by the user's optic nerve in the virtual reality scene is close to the real scene, this dimension is strongly related to video quality and is also part of immersion. Usually, we sample the duration and frequency of freezes in the system to represent how smooth the video information is. We have an intuitive performance indicator representing the stuttering situation and the frame rate. In addition, we also evaluate the black border ratio of the video, which is the proportion of black borders in the re-rendered picture when the head is turned in the scene.
    \item \textit{Interactive experience.} Interaction is a further requirement than viewing. In this dimension, we must not only consider the smoothness of video output. It is also necessary to evaluate whether the system's timely feedback when the user interacts. In addition to the subjective experience, we can abstract some performance indicators to represent the interaction quality. The more commonly used evaluation indicators are the number of GPU drawing calls. On the premise that the machine performance remains unchanged, reducing the number of GPU drawings can improve the efficiency of image drawing, thereby improving the smoothness of interaction. \cite{ware1994reaching}
\end{enumerate}

Based on the above evaluation ideas, we decided to evaluate the system performance based on the number of frames rendered and screen draw calls. Our benchmark platform is intel 8-core i7-8700 CPU, 16G memory, NVIDIA GeForce GTX 1080 discrete graphics, and the VR wearables used for testing are: HTC Vive head-mounted display device and accessories, HTC Vive infrared base station.
\section{Results}
\label{sec:results}
The frame rate test directly manifests the user's viewing experience, and rendering efficiency affects the frame rate. \cite{apteker1995video} In Unity's rendering engine, the CPU is responsible for preparing the model and related materials and textures and passing these data to the graphics processor for graphics rendering. This process is called \textit{DrawCall}. \cite{dickinson2015unity,dickinson2017unity} In large-scale projects, because the model properties in the scene are more complex, the number of \textit{DrawCalls} will be more. We consider using batch processing technology to reduce the number of draw calls per second. Batch processing means that objects of the same material can be processed simultaneously. In the Unity rendering engine, batch processing is divided into dynamic and static batch processing. In the scene, dynamic objects can be moved and transformed, and the rendering results of these objects will change with lighting and shadows. Unity treats all objects in the scene as dynamic objects by default and performs dynamic batch operations. However, in our system, there are a lot of static objects that will not be moved and deformed after rendering in the scene so that they can be rendered statically at a lower cost \cite{hocking2022unity}. In the system, we manually annotate these objects as static objects so that Unity can recognize them and batch them statically, thus reducing the number of draw calls per second. Here is the data we get after static batch optimization \cite{yin2019research,lan2022time}:

\begin{table}[htb] \centering
    \begin{tabular}{lcc} 
        \toprule
          & DrawCall/s(max) & DrawCall/s(avg) \\
        \midrule
        all dynamic load   & 880  & 796.346 \\
        with static load   & 795  & 663.618 \\
        \bottomrule
    \end{tabular}
    \caption{Draw call test}\label{table}
\end{table}

It can be seen from the test results that after static batch processing of some objects, the number of draw calls is significantly reduced, and a better frame rate performance is expected.

Before running the frame rate test, we need to introduce the concept of vertical sync. When a screen displays content, the pixels need to be refreshed from the horizontal direction, and each horizontal scan line is stacked in the vertical direction to form a complete picture. The vertical synchronization pulse signal is added between two frames of rendering. The graphics processor pauses and waits for the refresh time of the display after rendering a picture so that the output frame rate is consistent with the display's refresh rate. It can be seen from this that enabling vertical synchronization helps the graphics processor to synchronize the refresh rate of the display for graphics rendering, effectively avoids screen tearing, and can improve the user's viewing experience.  \cite{bourke2009idome} Therefore, we conducted frame rate tests with V-Sync turned on and off, respectively, and the results are as follows. \textit{f\_t} means frame time and  \textit{f\_r} means frame rate.

\begin{table}[!htb]
    \footnotesize
    \centering
    \renewcommand{\arraystretch}{1.0}
    \setlength\tabcolsep{2pt}
        \begin{tabular}{lcccc} 
            \toprule
             &  Max \textit{f\_t}(ms) & Avg \textit{f\_t}(ms) & Max \textit{f\_r}(times/ms) & Avg \textit{f\_r}(times/ms) \\
            \midrule
            V-Sync\_on   & 470.21  & 15.06 & 519.48  & 65.61\\
            V-Sync\_off   & 445.72  & 3.94 & 444.84  & 65.61\\
            \bottomrule
        \end{tabular}
    \caption{Frame rate Testing}\label{table}
\end{table}

It can be seen that the drop in frame rate after the vertical sync is turned on is in line with expectations. It is not a performance drop but a smoother screen refresh. The screen will be smoother in the real machine experience after turning on vertical sync.
\section{Discussion}
\label{sec:discussion}
This project developed a virtual reality based engine disassembly system collaborating with the company from a real user scenario. The system has a highly scalable software architecture and can be rapidly iterated through configurability in the face of new demand scenarios \cite{lan2018directed}. The performance test verified that the system can run smoothly on today's mainstream configuration machines. The high scalability and easy migration characteristics make an essential contribution to the system's popularity. 

With the development of virtual reality technology, its product form has gradually evolved from an immersive viewing experience that allows for human-computer interaction \cite{lan2019evolving}. With the constant maturity of the underlying technology, developers have quickly developed human-computer interaction programs through tools such as unity. In this context, developers can stimulate creativity and apply virtual reality technology in different scenarios, thus solving the challenges in these fields. On the other hand, our project applies advanced virtual reality technology to teach automotive engine repair. Through this technology, we hope to reduce the learning cost of these technicians, improve teaching efficiency, and further promote the development of the teaching model for this subject.

We believe that this teaching system will be deployed in many vocational institutions and soon contribute to improving vocational education quality. At the same time, we will further optimize the system in the future and strive for further improvement in interactive experience and performance.

\bibliographystyle{IEEEtran}
\bibliography{bibliography}

\end{document}